# Giant Optical Anisotropy in 2D Metal-Organic Chalcogenates

Bongjun Choi[1], Kiyoung Jo[1], Mahfujur Rahaman[1], Adam Alfieri[1], Jason Lynch[1], Greg K. Pribil[2], Hyeongjun Koh[3], Eric A. Stach[3,4], Deep Jariwala[1,3]*

[1]Department of Electrical and Systems Engineering, University of Pennsylvania, Philadelphia, Pennsylvania 19104, United States

[2]J.A. Woollam Co., Inc., 311 South 7th Street, Lincoln, NE 68508, United States

[3]Department of Materials Science and Engineering, University of Pennsylvania, Philadelphia, Pennsylvania 19104, United States

[4]Laboratory for Research on the Structure of Matter, University of Pennsylvania, Philadelphia, Pennsylvania 19104, United States

* Corresponding authors: dmj@seas.upenn.edu

**Abstract**

**Optical anisotropy is a fundamental attribute of some crystalline materials and is quantified via birefringence. A birefringent crystal not only gives rise to asymmetrical light propagation but also attenuation along two distinct polarizations, a phenomenon called linear dichroism (LD). Two-dimensional (2D) layered materials with high in- and out-of-plane anisotropy have garnered interest in this regard. Mithrene, a 2D metal-organic chalcogenate (MOCHA) compound, exhibits strong excitonic resonances due to its naturally occurring multi-quantum well (MQW) structure and in-plane anisotropic response in the blue wavelength (~400-500 nm) regime. The MQW structure and the large refractive indices of mithrene allow the hybridization of the excitons with photons to form self-hybridized exciton-polaritons in mithrene crystals with appropriate thicknesses. Here, we report the giant birefringence (~1.01) and tunable in-plane anisotropic response of mithrene, which stem from its low symmetry crystal structure and unique excitonic properties. We show that the LD in mithrene can be tuned by leveraging the anisotropic exciton-polariton formation via the cavity coupling effect exhibiting giant in-plane LD (~77.1%) at room temperature. Our results indicate that mithrene is an ideal polaritonic birefringent material for polarization-sensitive nanophotonic applications in the short wavelength regime.**

Introduction

Optical anisotropy, referring to the non-uniform response to light in crystals, is important for polarization-sensitive optical components and devices. One of the primary reasons for the optical anisotropic response is birefringence (Δn), which emerges due to variations in refractive indices of materials contingent upon the polarization of light and direction of propagation. Generally, optical anisotropy can be categorized into two distinct types, especially in two-dimensional (2D) crystals: out-of-plane anisotropy and in-plane anisotropy. For example, the transition metal dichalcogenides (TMDCs) of Mo and W exhibit a giant out-of-plane optical anisotropy owing to their inherent anisotropic layered van der Waals (vdW) structure[1,2]. For in-plane anisotropy, the crystals must have asymmetry along two axes in terms of crystal structure or electronic order in an in-plane direction. Several 2D crystals have also been shown to possess in-plane optical anisotropy arising from both structural and electronic or magnetic anisotropy[3-9]. Linear dichroism (LD), the difference in normal reflectance between incident light of perpendicular polarizations, is an effective tool[3,9] for probing in-plane optical anisotropy quantitively. Various 2D crystals such as black phosphorus (BP) and rhenium disulfide ($ReS_2$) have shown LD in the range of 20% to 40% at specific wavelengths[6,7], but further LD enhancement and spectral engineering are crucial goals that remain to be achieved. One approach to controlling LD is leveraging spin-charge coupling in antiferromagnet material or via nanopatterned plasmonic metasurfaces[3,10], but these approaches require cryogenic temperatures or extra lithographic steps. Moreover, though engineering of in-plane anisotropy is vital for photonic components and devices in the visible range, to the best of our knowledge no significant effort exists on engineering LD in the short wavelength regime (around 400-500 nm).

Mithrene has received the most attention among metal-organic chalcogenate (MOCHA) compounds due to its blue excitonic photoluminescence (PL) and large exciton binding energy ($E_b$) of ~ 400 meV[11-14]. Furthermore, the covalent bond between the Ag-Se atoms makes mithrene chemically stable under ambient conditions presenting a great advantage for optoelectronic applications[15] in contrast with other hybrid organic-inorganic semiconductors. The low symmetry of crystals in the in-plane direction provides anisotropic excitons[12,16] and is the fundamental requirement for achieving LD response. Furthermore, the MQW structure combined with the relatively high refractive index of mithrene results in strong light-matter interactions, leading to the formation of exciton-polariton without an external optical cavity known as self-hybridization[17]. This strong light-matter interaction creates new hybrid states referred to as the upper exciton-polariton (UEP) and lower exciton-polariton (LEP), distinguished by their respective higher and lower energy levels[18]. The low symmetric crystal structure and strong light-matter interaction properties of mithrene offer an unprecedented opportunity for engineering LD. In this paper, we present the giant birefringence and tunable in-plane LD in the mithrene leveraging the optical cavity to facilitate robust light-matter interaction and demonstrate the spectral tuning, particularly in the short wavelength regime (400 – 500 nm) without the use of any lithography process. Our findings indicate that mithrene is a promising polaritonic birefringent material for the amplification and tuning of LD for multi-spectral nanophotonic devices as a result of exciton-polariton formation.

**Results and Discussion**

To investigate the fundamental properties of mithrene and demonstrate enhanced LD from it, we synthesize mm-sized mithrene crystals using an organic single-phase reaction[15]. The large crystals are transferred on various substrates using mechanical exfoliation. The flake thickness ranges from a few-nm to a few hundred-nm, allowing us to study their thickness-dependent optical response and observe the impact of self-hybridization. Mithrene crystals belong to the monoclinic space group *P2₁/c* or *C2/c*, and the covalently bonded 2D Ag-Se layer is sandwiched by two organic phenyl rings, yielding a thickness of 1.4 nm per monolayer (Figure 1a)[15,19]. The top view of the in-plane lattice of the Ag-Se layer reveals a distorted hexagonal lattice, resulting in the in-plane anisotropic optical response. The isolation of each 2D Ag-Se layer with two organic ligands creates the bulk MQW structure, which is an efficient way to confine the excitons, facilitating the inherently strong light-matter interaction in mithrene crystals (Figure 1b). For sufficiently thick crystals of mithrene, this strong light-matter interaction forms self-hybridized bosonic quasi-particles and creates new hybrid polaritonic states called the LEP and UEP[20]. Further, the anisotropy of the crystal results in anisotropic exciton resonances, and interactions between the cavity photon and the excitons are also inherently anisotropic as shown in the schematic in Figure 1c.

We synthesize high-crystalline mithrene at mm$^2$ scale. Incorporation of long-chain amines, such as propylamine (PrNH$_2$), into the organic single-phase reaction facilitates the creation of silver-amine complexes during the synthesis process (detailed information can be found in Methods). This effectively retards crystal nucleation in the solution, thereby enabling the production of large-size, high-quality crystals[15,21]. With the use of long-chain amines, combined with optimized solution concentration and synthesis temperature, we produce mm-sized mithrene crystals (Figure 1d). A representative transferred crystal from the solution onto a substrate shows a 4.1 mm size scale which is large enough for fundamental studies and specific optoelectronic device applications (Figure 1e). The large size of the crystals also facilitates facile mechanical exfoliation using scotch tape or a Polydimethylsiloxane (PDMS) stamp due to their vdW structure nature[22]. As illustrated in Figure 1f, mithrene transferred onto a silicon dioxide/silicon (SiO$_2$/Si) substrate exhibits a large range of thicknesses, spanning from a few tens of nanometers to several hundred nanometers. This variance arises from the use of large-sized parent flakes/crystals. The diverse colors observed for varying flake thickness indicate the pronounced optical interference effects at play (Figure 1f). Mithrene crystals of varied thickness are transferred onto various substrates: SiO$_2$/Si, template-stripped gold (Au), aluminum oxide (Al$_2$O$_3$), and PDMS stamp substrates (Supporting Information (SI) Figure S1). The electron diffraction pattern obtained from the exfoliated mithrene shows a clear single-crystalline domain with sharp diffraction spots (SI Figure S2). The mithrene crystals on template-stripped Au are measured using Kelvin probe force microscopy (KPFM), showing high crystalline quality (SI Figure S3). Since mithrene has a direct bandgap at the Γ-point with a simple parabolic band edge[13], as-synthesized mithrene crystals demonstrate distinct blue PL at approximately 467 nm (2.65 eV) at room temperature. The PL peak is further characterized by an exceedingly narrow full width at half maximum (FWHM) of approximately 17.1 nm (0.10 eV). Upon cooling to 80K, the peak blue shifts to 454 nm (2.73 eV) while concurrently displaying an even narrower FWHM of about 4.5 nm (0.03 eV) (Figure 1g). We note that the intrinsic PL emission energies of thin (4 layers) vs. thick mithrene flakes do not change as a function of thickness, and therefore they exhibit no thickness-dependent quantum confinement effects (SI Figure S4), unlike TMDCs of Mo and W. Further, the temperature-dependent PL from 80K to 300K indicates gradual blue-shifts with decreasing temperature, and no clear signatures of trap state emission, suggesting the high opto-electronic quality of the crystals (SI Figure S5). Finally, we perform Raman spectroscopy to structurally characterize the mithrene crystals and observe the low energy phonon modes that tentatively correspond to intrinsic [AgSePh]$_\infty$ modes (Figure 1h)[16]. In addition, Raman peaks resulting from vibrations, rotations, and stretching of the organic ligands are also identified above 200 cm$^{-1}$. The absence of the δ(CseH) scissoring mode which originated from neat benzeneselenol (PhSeH) at 796 cm$^{-1}$ signifies the exclusive presence of silver-bonded PhSe species, indicating the high quality of the synthesized mithrene[23,24].

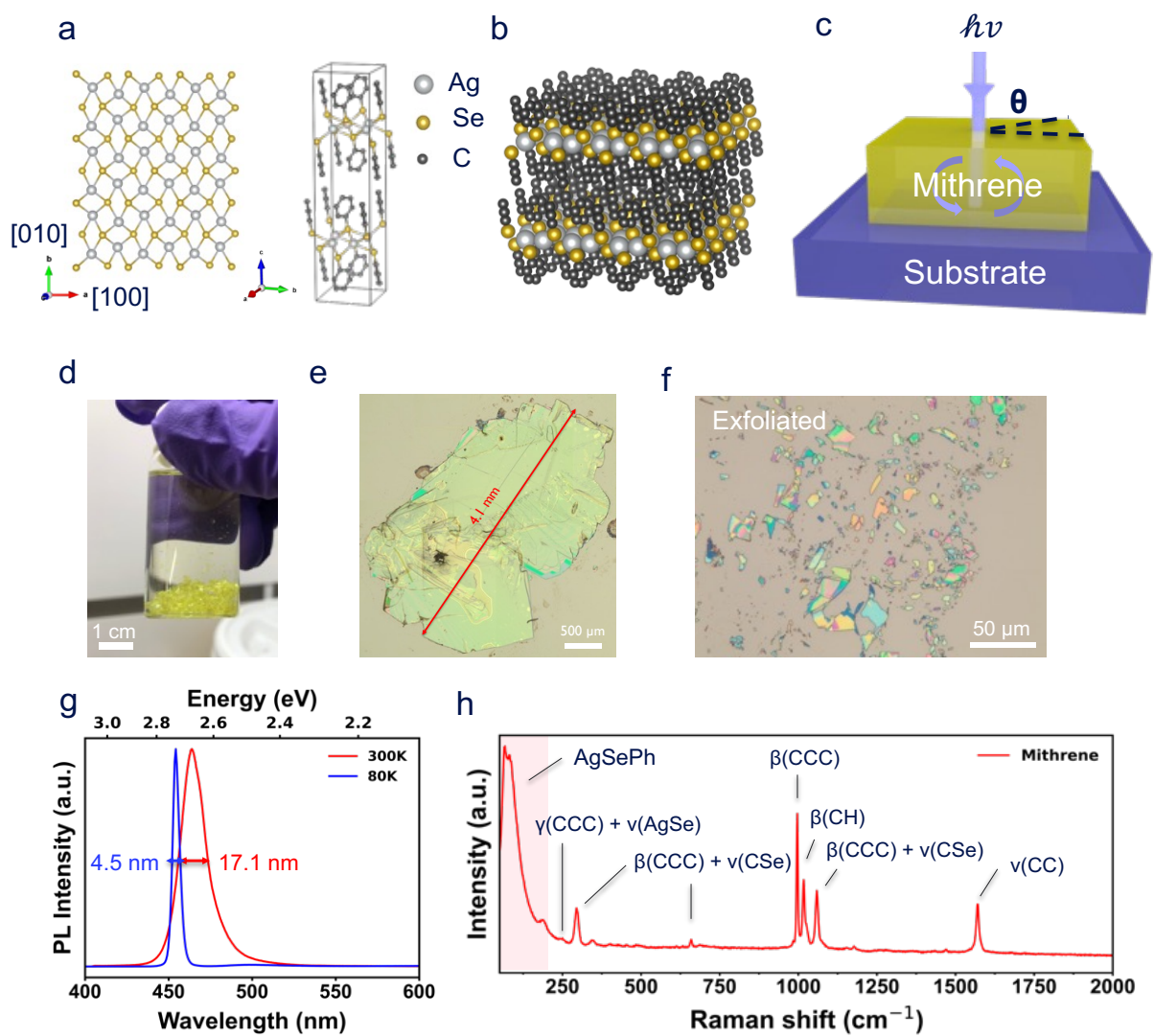

**Figure 1. Large single crystalline mithrene and its optical properties.** (a) Schematic of the unit cell of mithrene (right) and top view (left) of the Ag-Se bonds in-plane in mithrene. (b) Schematic of MQW structure with Ag-Se covalently bonded planes sandwiched between organic phenyl groups. (c) Schematic of in-plane anisotropic light-matter interactions mithrene crystals on the arbitrary substrate. Confined cavity photons in the crystals interact with the excitons, and the interaction is dependent on the polarization (θ) of incidence light due to the anisotropic nature of the excitons. (d) Synthesized mithrene crystals in the reaction vial showing numerous yellow-colored, translucent mm-sized mithrene crystals. (e) Transferred large mithrene crystal (~4.1 mm) on a SiO$_2$/Si substrate using the drop-casting method. The image was constructed by stitching together multiple pictures acquired using an optical microscope (5x objective). (f) Mechanically exfoliated mithrene using scotch tape or PDMS on the SiO$_2$/Si substrate, indicating various sizes and thicknesses of exfoliated mithrene. (g) PL spectra of mithrene at 300K and 80K with 405 nm excitation, respectively. (h) Raman spectrum of mithrene at room temperature with 633 nm excitation with the appropriately labeled peaks. Mithrene exhibits low energy phonon modes at 62 cm$^{-1}$, 87 cm$^{-1}$, 192 cm$^{-1}$ as well as phenyl vibration mode and pure AgSe modes at 249 cm$^{-1}$, 297 cm$^{-1}$, 661 cm$^{-1}$, 998 cm$^{-1}$, 1018 cm$^{-1}$, 1060 cm$^{-1}$, and 1572 cm$^{-1}$.

To investigate and quantify the giant optical anisotropy in mithrene crystals, we first perform Mueller matrix ellipsometry to calculate the three-dimensional dielectric tensor of mithrene. The Stokes vector provides a complete description of the amplitude, phase, and polarization of light. The Mueller matrix describes the relation between incident and reflected light's Stokes vector, providing a complete description of the light-matter interaction[25]. The refractive indices along two different optical axes ([100] and [010]) are then extracted using the multi-Tauc-Lorentz oscillator model from the Mueller matrix (SI Figure S6)[26]. We also extract the refractive index of mithrene along the out-of-plane ([001]) from the Mueller matrix. As expected due to its layered structure, mithrene exhibits clear anisotropy along the out-of-plane direction. In addition, the extinction coefficient along the [010] direction, which is related to the absorption of materials, exhibits two dominant exciton transitions ($X_1$ and $X_3$ exciton absorptions), while along the [100] direction one dominant excitonic transition ($X_2$) is observed (Figure 2a), which are well aligned with the observation in previous studies[12]. Mithrene displays significant in-plane birefringence ($\Delta n = n_{[010]} - n_{[100]}$) in the 400-500 nm range from -0.31 to 1.01 as a function of wavelength, with a particularly notable value of 1.01 at 440 nm. Likewise, the out-of-plane birefringence ($\Delta n = n_{[001]} - n_{[010]}$) also shows a large value at the short wavelength regime ranging from -0.74 to 0.65 (Figure 2b). We identify three prominent peaks in the in-plane birefringence spectrum. It's worth noting that there exists a close relationship between the refractive index and the extinction coefficient, as governed by the Kramers-Kronig relation; consequently, this connection extends to optical anisotropy. Therefore, the birefringence spectrum shows a close relation to the LD spectrum of the weak coupling configuration discussed later. We systematically compare the absolute value of birefringence of mithrene with other well-known birefringent materials such as hexagonal boron nitride (hBN), barium titanate ($BaTiO_3$), calcite ($CaCO_3$), BP, $ReS_2$, and $BaTiS_3$ using experimental and calculated in- and out-of-plane birefringence. Mithrene indicates giant in-plane birefringence (~1.01) in the short wavelength regime (400-500 nm) compared to other in-plane anisotropic materials (Figure 2c). To the best of our knowledge, mithrene exhibits the highest in-plane birefringence within the short, visible wavelength regime when compared to other known materials. In addition, mithrene exhibits substantial birefringence in both the in-plane and out-of-plane directions, and these exceptional properties of mithrene make it a highly promising anisotropic material for various photonic applications.

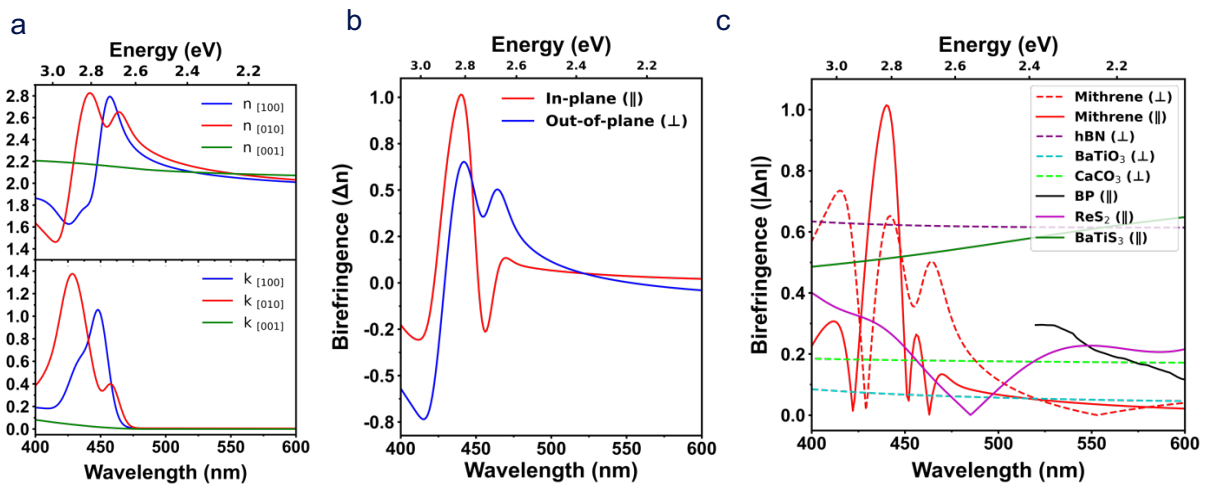

**Figure 2. Refractive index along the different axes and in- and out-of-plane birefringence of mithrene.** (a) In- and out-of-plane refractive indices and extinction

coefficient along the different axes. (b) In- and out-of-plane birefringence of mithrene. (c) A comparison of the magnitude of in- and out-of-plane birefringence as a function of wavelength between various in-plane anisotropic materials from previous literature[27-32] (Plot with molybdenum disulfide ($MoS_2$) can be found in SI Figure S7). The symbols of ∥ (solid line) and ⊥ (dotted line) stand for in- and out-of-plane, respectively.

In our investigation of optical anisotropic response in mithrene crystals, we perform polarization-dependent PL and reflectance measurements. We control the incident light's polarization (405 nm laser for PL and white light source for reflectance measurement) by inserting a polarizer and the half-waveplate (HWP) between the light source and sample, and the signals, contingent on polarization, are collected by the charge-coupled device (CCD) detector (Figure 3a). We initially record the PL spectrum by rotating the HWP (θ) for polarization control. Each recorded spectrum is characterized by a unique PL intensity, which originates from different exciton excitations depending on the polarization of incident light. Mithrene exhibits three main excitonic transitions ($X_1$, $X_2$, and $X_3$), and each exciton can be excited by specific polarization of light[12]. The PL peak is observed at 467 nm (2.65 eV) with polarization-dependent intensities at room temperature, which corresponds to the lowest-energy exciton ($X_1$) (Figure 3b). The peak intensity of each spectrum is plotted in polar coordinates and fitted using the formula A = ($A_{max}$ − $A_{min}$) $\cos 2(θ − θ_0)$ + $A_{min}$, where $θ_0$ represents the reference polarization angle corresponding to the point at which the PL reaches its maximum intensity (Figure 3c)[33]. Our observations unveil clear in-plane anisotropy, which is discerned both in the experimental data points and in the results of the fitted line. Furthermore, the primary polarization axes ([010] and [100] direction) are identified through polar plots since the $X_1$/$X_3$ excitons are strongly associated with the transition dipole moments along the [010] direction (Ag-Ag chain direction) while $X_2$ exciton is related with the [100] direction (Se-Se chain direction)[12]. To visualize and substantiate this behavior, we further examine the mithrene flakes using polarized optical microscopy. The significant change in color with a polarizer rotation angle ranging from 0 to 180 degrees indicates the clear in-plane anisotropy behavior of mithrene (SI Figure S8). To quantify the in-plane anisotropy, we perform polarization-dependent reflectance measurements and calculate the LD of mithrene. LD is defined as $\frac{R_{[010]} - R_{[100]}}{R_{[010]} + R_{[100]}}$, where $R_{[010]}$ ($R_{[100]}$) corresponds to the intensities of vertically (horizontally) polarized optical reflection, respectively. Since the cavity effect modulates the LD intensity, we initially investigate the weak coupling regime by using a thin mithrene flake on the $SiO_2$ (100 nm)/Si substrate to probe the LD properties of mithrene itself. In Figure 3d, the optical microscopy and the atomic force microscopy (AFM) image illustrate the corresponding sample configuration, and the white line cut in the AFM image indicates the 13.7 nm thickness of mithrene. The transfer matrix method (TMM) simulation results indicate that the configuration does not show strong cavity-enhanced absorption (SI Figure S9). We conduct the polarization-dependent reflectance and plot two representative reflectance spectra along the [010] and [100] directions, revealing different responses (Figure 3e). The polarization-dependent reflectance spectrum indicates three dominant reflectance dips (absorption peaks) corresponding to the $X_1$, $X_2$, and $X_3$ excitons with good agreement with previous observations[12] and mithrene's extinction coefficients in Figure 2a. In the [010] direction, strong absorption peaks attributed to the $X_1$ and $X_3$ excitons are observed, while the $X_2$ absorption dominates along the [100] direction. Polarization between the [010] and [100] directions shows the combination of reflectance dips from the $X_1$, $X_2$, and $X_3$ excitons (SI Figure S10). Furthermore, the orthogonal behavior of excitons between $X_1/X_3$, and $X_2$ is also observed (SI Figure S11), and this observation aligns well with findings from prior studies[12]. In conditions where the cavity effects are minimized, mithrene crystals exhibit ~24.4% of the LD magnitude with the three dominant peaks. These LD peaks are linked to exciton transitions and are responsible for the observed high LD magnitude, as illustrated in Figure 3f. Therefore, the LD spectrum of the weak coupling configuration shows a close relation to the birefringence spectrum in terms of the peak position and the trend since both are highly related to the excitonic resonance (SI Figure S12). The distinctive spectral features arising from these exciton transitions play a pivotal role in shaping the LD response within the mithrene crystals. Moreover, this observation assumes that 24.4% of the LD magnitude is attributed to the inherent low symmetry of crystals in the in-plane directions. Next, for the LD amplification with the optical cavity, we deliberately positioned the thick mithrene crystals (~184 nm) on the $Al_2O_3$ (10 nm)/Ag (100 nm) substrate to promote the large phase shift and back reflection for leveraging the Fabry-Pérot cavity resonance mode (Figure 3g). The thick mithrene refers to a few hundred nanometers to attain the phase shift of 180º to satisfy phase matching conditions for the cavity mode[34], and the vdW nature of mithrene enables the facile thickness control of crystals. This configuration ensures a substantial phase shift, thus enabling constructive interference in the optical cavity[35]. Interestingly, mithrene forms an exciton-polariton without an external cavity due to its MQW structures and high refractive index, enabling near-unity absorption. These self-hybridized exciton-polariton formations lead to strong absorption modes that are the UEP (●), middle exciton-polariton (MEP) (★), and LEP(■), and two representative spectra along [010] and [100] directions indicate the still anisotropic response of mithrene crystals due to anisotropic light-matter interactions (Figure 3h). Since two excitons ($X_1/X_3$) are dominant along the [010] direction, three reflectance dips can be found including MEP (★) along the same direction. We find that the reflectance dips corresponding to the LEP resonance around 460 nm are different depending on the incident polarization of the light ([010] and [100]), resulting in a giant magnitude of LD response (~77.1%). This manipulation increases LD magnitude approximately 3.2 times compared to the absence of the optical cavity at room temperature (Figure 3i). It is worth noting that this increase in LD intensity primarily originated from mithrene crystals due to light-matter interaction since the Ag back reflector barely absorbs in the visible regime (SI Figure S13). With regard to the peak position of LD, our observations indicate a change in the LD peak positions compared to Figure 3f. This change is attributed to the exciton-polariton hybridization effect, which results in the formation of new anisotropic hybrid states, UEP and LEP, along different optical axes. These new hybridized states play a crucial role in modifying the energy landscape, which affects the peak positions in the LD spectrum, implying that exciton-photon hybridization can be an unprecedented strategy for tuning the LD response of excitonic materials.

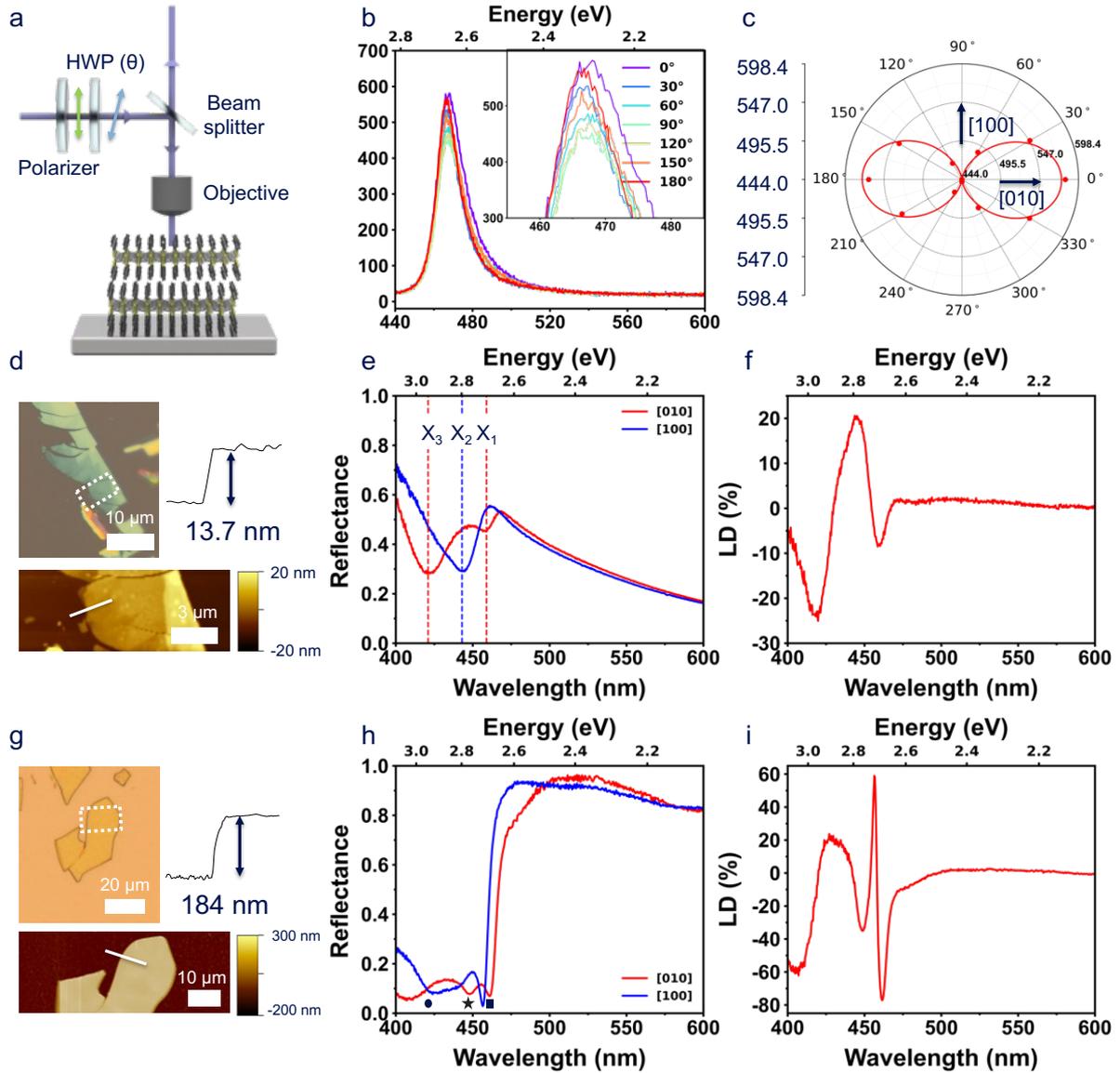

**Figure 3. In-plane optical anisotropy and linear dichroism of mithrene with and without an optical cavity.** (a) Schematic of the experimental setup for the polarization-dependent PL and reflectance. (b) Polarization-dependent PL measurement at room temperature. (c) The polar plot at the excitonic peak (~ 467 nm) shows the clear in-plane anisotropic behavior. (d) Optical microscopy and AFM image of minimized cavity effect configuration, which is the thin mithrene (~13.7 nm) on the SiO$_2$ (100 nm)/Si substrate. (e) Polarization-dependent reflectance measurement of thin mithrene on the SiO$_2$/Si substrate. (f) The magnitude of LD as a function of wavelength shows a large LD magnitude of around 24%. (g) Optical microscopy and AFM image of thick mithrene (~ 184 nm) on the Al$_2$O$_3$ (10 nm)/Ag (100 nm) substrate for the cavity resonance mode (h) Polarization-dependent reflectance measurement at room temperature of thick mithrene on the Al$_2$O$_3$/Ag substrate. The blue and red marked regions denote the UEP and LEP peaks respectively. (i) The LD value of mithrene as a function of wavelength indicates a giant LD corresponding to 77.1%.

For a deep understanding of the anisotropic exciton-photon interactions without the external cavity, we simulate the reflectance as a function of mithrene's thickness using the TMM calculations[36,37] and calculate the reflectance depending on the crystal orientation along [010] and [100] directions. The reflectance dips in experimental data are plotted as a black star on the calculated data as a function of the mithrene thickness (Figure 4a, b). Both the simulated and experimental results manifest pronounced self-hybridization, as evidenced by the distinct anti-crossing behavior in the calculated reflectance spectra for both polarization axes. The resultant polariton branches are then fit using the two ([100] direction) and three ([010] direction)-coupled oscillator mode[35]. Since mithrene shows two dominant excitons ($X_3$ and $X_1$) along the [010] direction, the MEP, which is fitted by the three-oscillator model, can be found along the [010] direction. The Rabi splitting ($\hbar\Omega$) evaluated from three and two-coupled oscillator models indicates large splitting energy $\hbar\Omega$ = 586 and 316 meV ($X_3$ and $X_1$ excitons along [010]) and 587 meV ($X_2$ exciton along [100]), respectively. These strong coupling strengths along two axes between the exciton mode and the cavity photon mode ensure robust anisotropic light-matter interactions. Figure 4c shows the simulated reflectance spectrum for flakes of varying thickness and along polarization axes ([010] and [100]). Clear differences between the two polarization axes are readily discernible, with a distinct observation of anisotropic UEP (●) and LEP (■) branches evident in each varying thickness of the material. Since the cavity photon energy is influenced by the cavity length, the hybridized states also vary depending on the thickness of mithrene which acts as a self-cavity. The corresponding experimental reflectance spectrum indicates a close qualitative match in terms of the in-plane anisotropic behavior and peak position (Figure 4d). Note that the presence of UEP below 400 nm remained undetectable in our observation, due to the wavelength resolution limitations of our CCD detector. In addition, to further verify the anisotropic behavior in the PL emission, we measure the PL and reflectance at a low temperature (80K). The reflectance spectrum at the low temperature indicates the position of the UEP (●) and LEP (■) states. A higher order lower polariton (HO) (★) branch can be seen, which originates from the interaction between the exciton and the higher order Fabry-Pérot cavity modes in mithrene (SI Figure S14). The PL spectrum shows two dominant emission peaks at 453 nm and 510 nm, corresponding to HO and LEP mode, respectively. We further measure the polarization-dependent PL at a low temperature (80K) and observe the two peaks from HO and LEP by varying polarization angles (Figure 4e). The PL spectrum of mithrene exhibits anisotropic behavior, as detected through variations in PL intensity corresponding to changes in the polarization angle. The angular dependence of both HO and LEP emissions is shown in Figure 4f, indicating clear in-plane anisotropy in both HO and LEP mode (LEP peak intensity is multiplied by 15 for the plot). The observations from polarization-dependent reflectance and PL measurement affirm the existence of self-hybridized exciton-polaritons with in-plane anisotropy in the mithrene, playing an important role in tuning the LD properties.

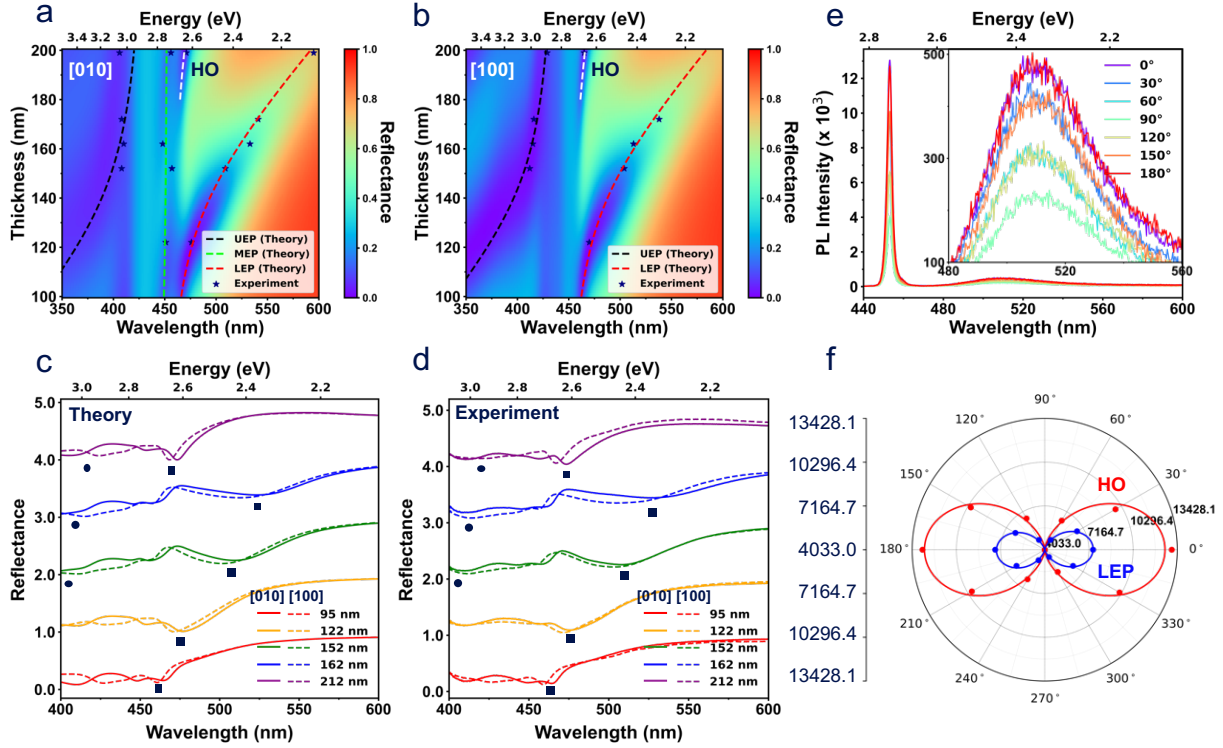

**Figure 4. In-plane anisotropic exciton-polariton formation.** (a) The simulated reflectance spectra as a function of the mithrene thickness on template-stripped Au substrate with UEP, MEP, and LEP fitted by three and two-coupled oscillator models along [010] (b) and [100] directions. The white dashed line stands for the HO mode in each direction. (c) The calculated (d) and experimental reflectance spectrum with UEP (●) and LEP (■) markers along two different polarization axes varying the mithrene thickness indicate anisotropic exciton-polariton formation demonstrating a high degree of consistency between simulated and experimental data. (e) Polarization-dependent PL measurement at 80K with the sample in SI Figure S14, showing anisotropic behavior not only in HO mode but LEP mode due to the self-hybridized in-plane anisotropic exciton-polariton formation. (f) The polar plot at the HO and LEP (x15 magnified) peak indicates clear in-plane anisotropic behavior.

The self-hybridization in mithrene allows not only the enhancement of the LD intensity but also the tunability of spectral peak position due to the creation of new UEP and LEP states. To demonstrate the spectral tuning, we study the three different configurations (1. thin mithrene on $SiO_2$/Si, 2. thin, and 3. thick mithrene on template-stripped Au) to identify the impact of anisotropic exciton-polariton formation on the LD peak position (Figure 5a). The thin mithrene sample is too thin to sustain a Fabry-Pérot resonance mode on either the $SiO_2$/Si or Au substrate, so no strong coupling occurs. Therefore, the LD peak position in the thin mithrene sample is highly related to the birefringence peak position, which stems from the mithrene's excitonic features ($X_1$, $X_2$, and $X_3$). Since thin mithrene is not appropriate for the optical self-cavity regardless of the substrate, consequently, both configurations 1 and 2 indicate similar LD trends (Figure 5a). The simulated LD spectra, which are obtained from two configurations (1 and 2), substantiate the experimental observation. In contrast, thick mithrene on template-stripped Au (configuration 3), which can support self-hybridized exciton-polaritons, exhibits a distinct behavior regarding the peak position in the LD spectrum. In configuration 3, our observations reveal the presence of two distinct peaks (UEP and LEP) within the LD spectrum. This behavior is attributed to the formation of in-plane anisotropic exciton-polaritons, and the UEP (LEP) peak position is blue (red)-shifted owing to hybridized states. Using this anisotropic hybridization, the LD peak position can be tuned approximately from 400 to 500 nm with a high magnitude of LD. We simulate the LD as a function of the thickness of mithrene (Figure 5b). The black star marker (★) on simulated data corresponds to experimental LD peak data collected from various thicknesses of mithrene. The LD peak (either red or blue) is well-matched with the polaritonic branches since the hybridized polariton state determines the LD spectral position and gives tunability of LD along the polaritonic branches. This observation serves to validate that the tunability of the LD response can be achieved by altering the thickness of the optical cavity.

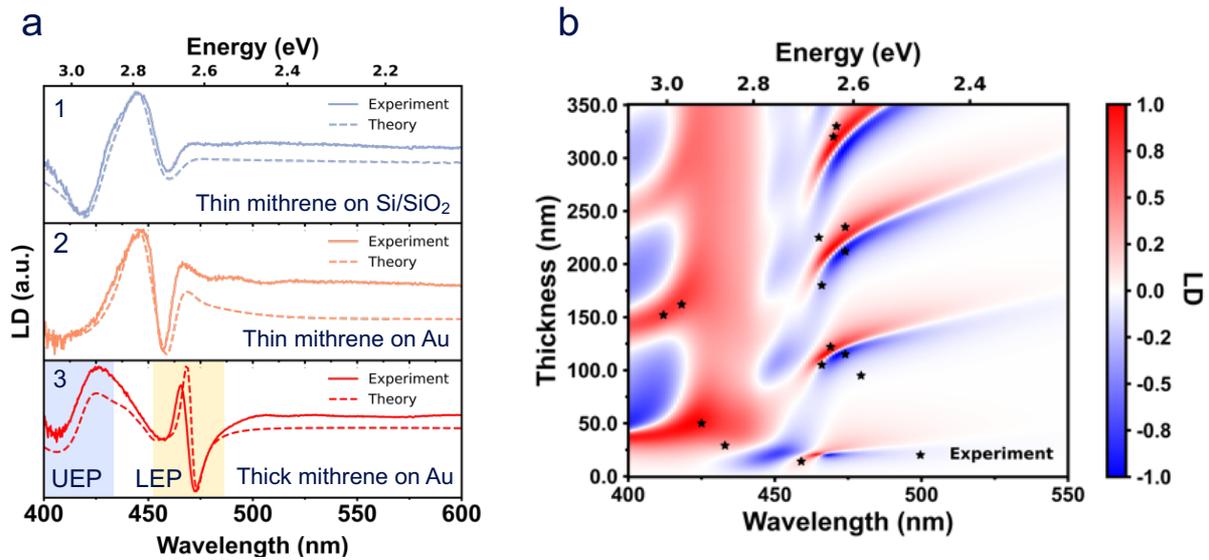

**Figure 5. LD spectral response tuning.** (a) Experimental and theoretical LD peak position depending on the configuration (1. thin mithrene on the $SiO_2$/Si, 2. thin, and 3. thick mithrene on template-stripped Au). The new LD peak position emerges with the thick mithrene on the Au substrate due to anisotropic exciton-polaritons formation. (b) The simulated LD spectrum as a function of mithrene thickness on template-stripped Au with experimental LD peak position (★).

## Conclusion

In conclusion, we observe the giant birefringence and LD which is tunable in terms of both intensity and spectral response by leveraging the exciton and photon hybridization with the optical cavity in mithrene crystals. We synthesize high-quality and large mithrene crystals using an organic single-phase reaction, and then transfer the crystals to the various substrates with the mechanical exfoliation for this study. A significant in-plane birefringence (Δn) of 1.01 is detected at 440 nm and large out-of-plane birefringence around -0.74 at 415 nm due to its inherent vdW structure of mithrene. The in-plane anisotropy of mithrene is further investigated with polarization-dependent PL and reflectance measurements, exhibiting a large magnitude of LD corresponding to 24.4% without cavity enhancement. The magnitude of LD is amplified by leveraging optimized optical cavity systems, resulting in a 3.2-fold enhancement to approximately 77.1% at room temperature. The optical cavity allows spectral tuning of LD ranging from 400 nm to 500 nm. Our investigation reveals that the in-plane anisotropic exciton-polariton formation significantly influences both the magnitude and spectral tuning observed in our experimental results. This finding is corroborated by theoretical TMM calculations, demonstrating a high degree of alignment between experimental observations and theoretical calculation. Our findings indicate that exciton-polariton offers a means to achieve tunable in-plane optical anisotropic response, and mithrene is a promising polaritonic material for polarization-sensitive and multi-spectral nanophotonic devices due to both large in- and out-of-plane birefringence and tunable in-plane LD, especially in short wavelength regimes with a simple design.

**Methods**

**[AgSePh]$_\infty$ (Mithrene) synthesis.**

Mithrene crystals are synthesized by using the organic single-phase method with additional propylamine (PrNH$_2$)[15]. Silver nitrate (AgNO$_3$) is dissolved in a solution of PrNH$_2$ with a concentration of 10-15 mM, while diphenyl diselenide (Ph$_2$Se$_2$) is dissolved in toluene with the same concentration as the AgNO$_3$/PrNH$_2$ solution. These two precursor solutions are mixed with 50% v/v PrNH$_2$/Toluene and stored at room temperature for 5-7 days to form the large crystals.

**Sample preparation.**

As-synthesized mithrene in PrNH$_2$/Toluene solution is thoroughly washed with the isopropyl alcohol (IPA) and dispersed in IPA solution before drop casting on the arbitrary substrate. The dispersed mithrene crystal in IPA solution is dropcasted on the target substrate which is placed on a 80 °C hot plate and the residual solvent is quickly evaporated. The dropcasted mithrene is picked up with PDMS or scotch tape for mechanical exfoliation to control its thickness, and we transfer the flake to the target substrate. We mainly use the dry oxidized SiO$_2$/Si wafer and template-stripped Au with an extremely smooth surface of 0.5 nm root mean square (RMS) value[38]. The flake's thickness and surface roughness are investigated using AFM.

**Optical characterization (PL, reflectance, and temperature, polarization dependent).**

The PL, Raman, and reflectance spectrum are collected using the Horiba LabRam HR Evolution confocal microscope. For the PL measurement, we use a 405 nm continuous wave (CW) laser excitation and 100 grooves/mm grating for collecting the signal to the CCD detector. The polarization of incidence light is controlled by rotating polarized in front of the light sources. The Raman signal was recorded using 600 grooves/mm grating for better resolution with a 633 nm CW laser excitation. The reflectance is measured using the white light source (AvaLight-HAL), while controlling the polarization of incidence light with the polarizer. We assume that the silver mirror has a perfect reflection in the region of interest (400-800 nm) and use the formula below for normalization.

$$R_{normalized} = \frac{R_{Mithrene} - R_{background}}{R_{silver} - R_{background}}$$

The temperature-dependent measurement is conducted using the Linkam stage (THMS600). The sample is placed inside the Linkam stage, cooling down to 80K controlling the flow rate of liquid nitrogen.

**Optical simulation**

Theoretical reflectance, absorbance, and LD are calculated using the TMM calculation with homemade python code[36,37].

**Ellipsometry measurement.**

The dielectric function of mithrene is obtained using ellipsometry. Since the sample size is small and has the anisotropic dielectric function, we use the ellipsometer with the micro spot size (~ 100 μm) utilizing a focusing optical module and measured the Mueller matrix at an incidence angle of 65º. The dielectric function along the different axes is extracted from the Mueller matrix using the multi Tauc-Lorentz oscillator model with CompleteEase software from J.A. Woollam.

**KPFM.**

For the KPFM measurement, we use an OmegaScope-R (AIST-NT) setup. The Au-coated AFM probe (HQ:NSC14/Cr-Au) is used for KPFM measurement. The conductive tip (Au) is biased with 3V while the sample is grounded and scanned the region of interest, obtaining the topography map simultaneously.

**Acknowledgments.**

D. J., B.C. acknowledges support from the Office of Naval Research Young Investigator Award (N00014-23-1-203). D.J., A.A. and J.L. also acknowledge partial support from the Asian Office of Aerospace Research and Development of the Air Force Office of Scientific Research (AFOSR) (FA2386-21-1-4063). A.A. acknowledges partial support for the work from the Vagelos Institute of Energy Science and Technology Graduate fellowship. M.R. acknowledges support from Deutsche Forschungsgemeinschaft (DFG, German Research Foundation) for Walter Benjamin Fellowship (award no. RA 3646/1-1). This work was carried out in part at the Singh Center for Nanotechnology, which is supported by the NSF National Nanotechnology Coordinated Infrastructure Program under grant NNCI-2025608 and through the use of facilities supported by the University of Pennsylvania Materials Research Science and Engineering Center (MRSEC) DMR-2309043.

Supporting Information

# Giant Optical Anisotropy in 2D Metal-Organic Chalcogenates

Bongjun Choi[1], Kiyoung Jo[1], Mahfujur Rahaman[1], Adam Alfieri[1], Jason Lynch[1], Greg K. Pribil[2], Hyeongjun Koh[3], Eric A. Stach[3,4], Deep Jariwala[1,3*]

[1]Department of Electrical and Systems Engineering, University of Pennsylvania, Philadelphia, Pennsylvania 19104, United States

[2]J.A. Woollam Co., Inc., 311 South 7th Street, Lincoln, NE 68508, United States

[3]Department of Materials Science and Engineering, University of Pennsylvania, Philadelphia, Pennsylvania 19104, United States

[4]Laboratory for Research on the Structure of Matter, University of Pennsylvania, Philadelphia, Pennsylvania 19104, United States

* Corresponding authors: dmj@seas.upenn.edu

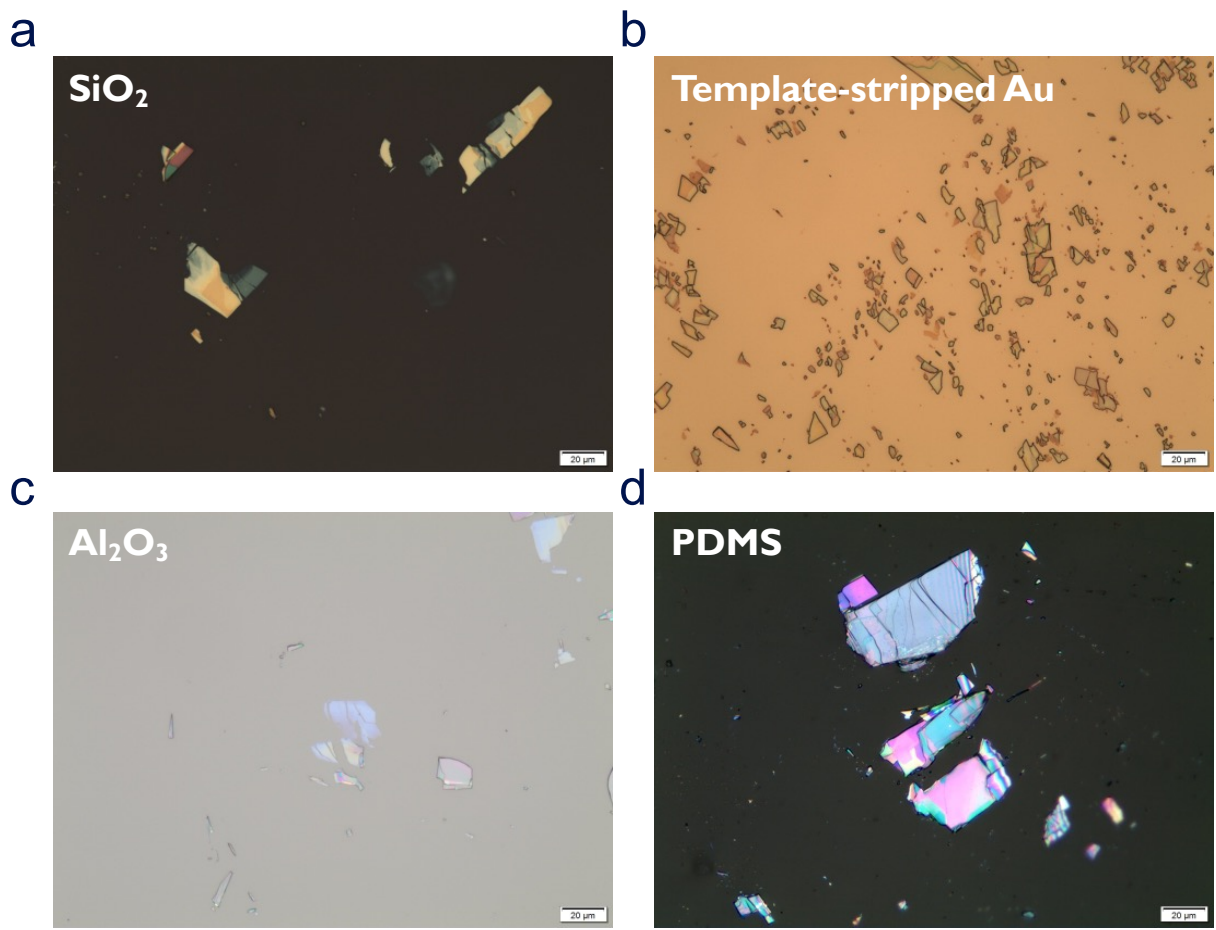

**Figure S1.** Optical microscopy (OM) image of transferred mithrene using mechanical exfoliation on the (a) Si/SiO$_2$, (b) Template-stripped gold (Au), (c) c-plane Al$_2$O$_3$, and (d) Polydimethylsiloxane (PDMS) substrate.

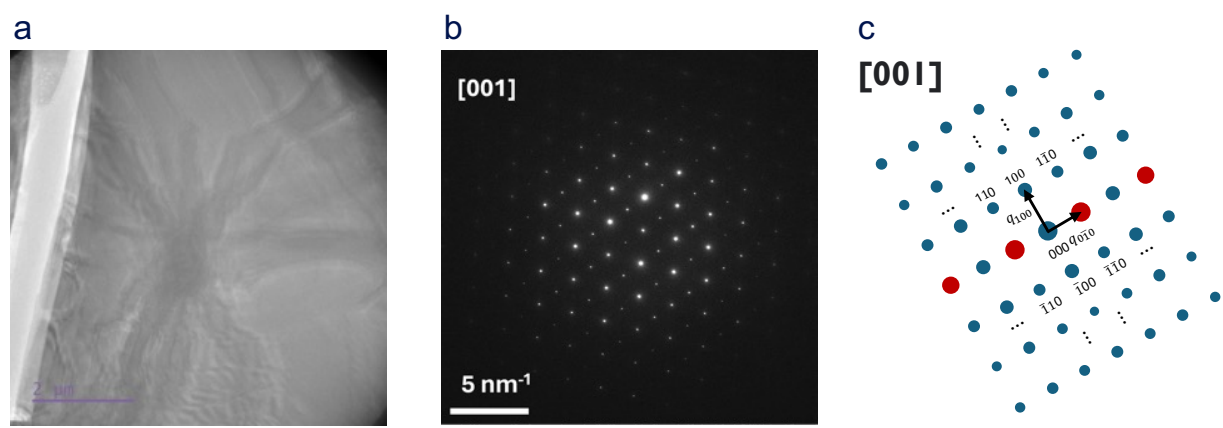

**Figure S2.** Transmission electron microscopy (TEM) analysis of the exfoliated mithrene. (a) Mithrene flake on the TEM grid shows clear Kikuchi lines indicating an overall single crystallinity of the exfoliated mithrene. (b) The electron diffraction pattern taken from [001] zone axis in Figure S2a exhibits a single crystalline domain that aligns with (c) its simulated patterns. The calculated lattice parameter from the diffraction patterns is identical to the reported values[1] (a=5.833, b= 7.29, β=95.58°). The red dots in Figure S2c denote double diffraction patterns that arise from multiple electron diffraction.

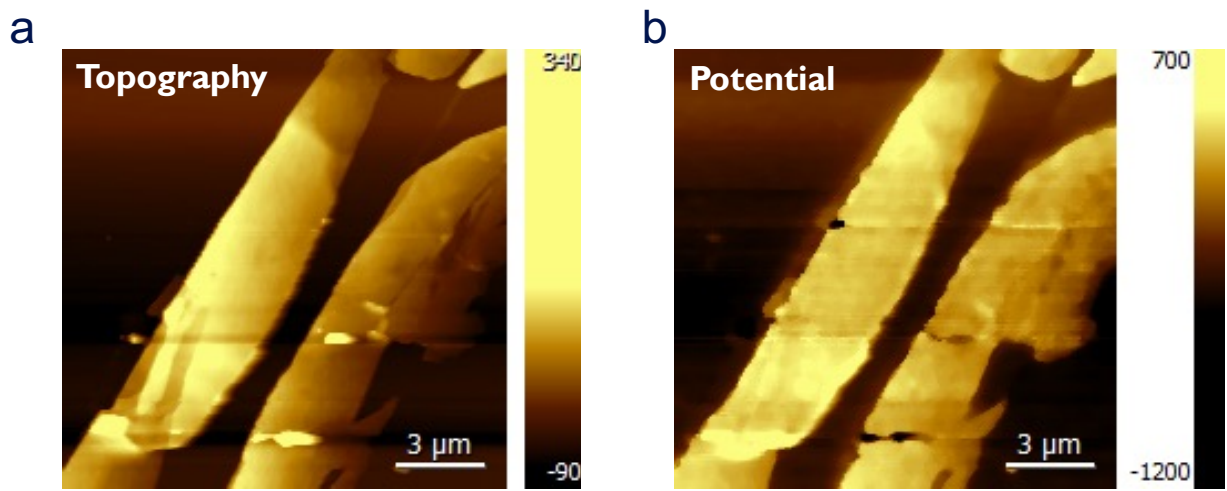

**Figure S3.** Comparison between the (a) topography and (b) surface potential map of mithrene on template-stripped Au obtained by atomic force microscopy (AFM) and kelvin probe force microscopy (KPFM), respectively.

The topography map reveals fluctuations attributed to variations in the height of transferred mithrene crystals. In contrast, the surface potential map exhibits uniform crystals, indicating the high crystallinity quality of the transferred flakes. This distinction between the topography and potential maps suggests that while the height of the crystals may vary, their surface potential remains consistent and indicative of a single crystal structure, as shown in the surface potential map.

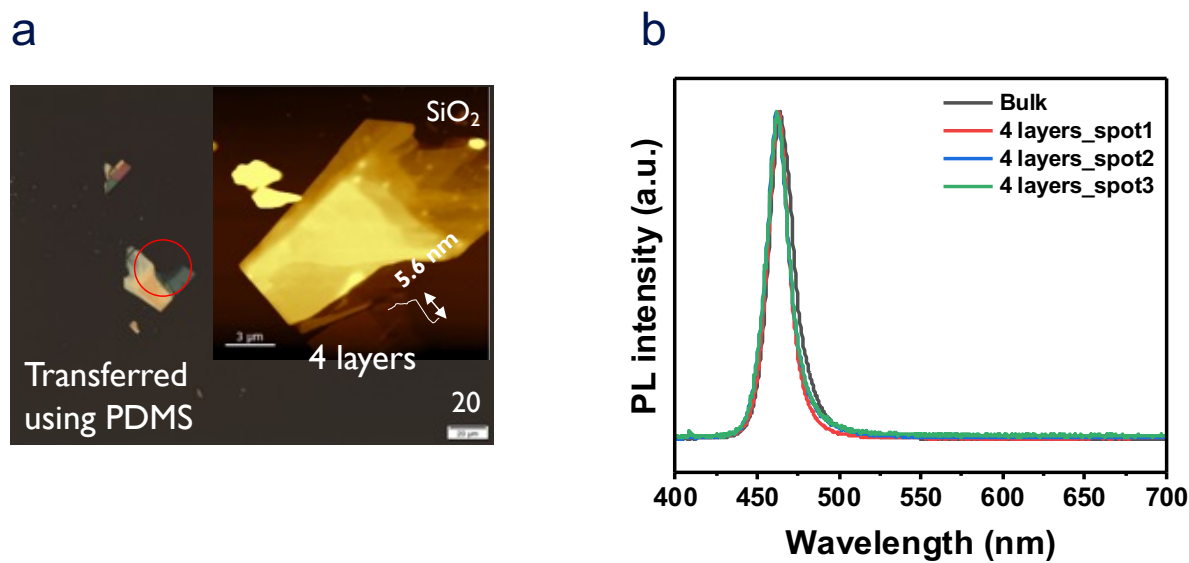

**Figure S4.** (a) OM and AFM images of transferred thin mithrene, indicating ~5.6 nm of thickness which corresponds to 4 layers of mithrene. (b) Photoluminescence (PL) measurement of bulk and thin mithrene crystals.

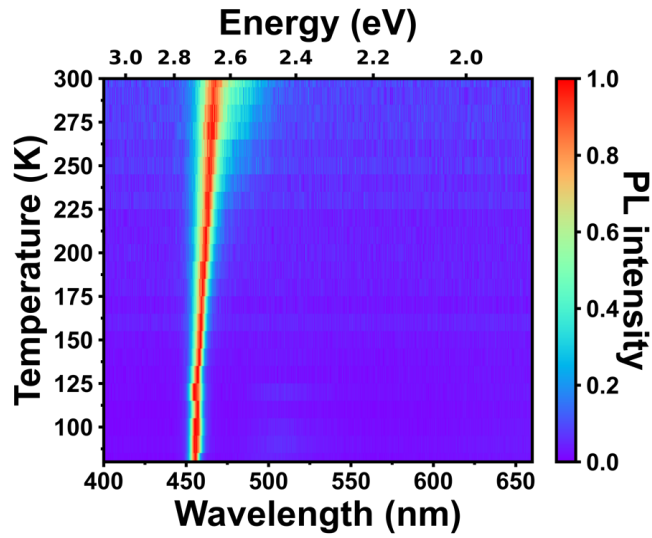

**Figure S5.** 2D color map of temperature-dependent PL from 80K to 300K.

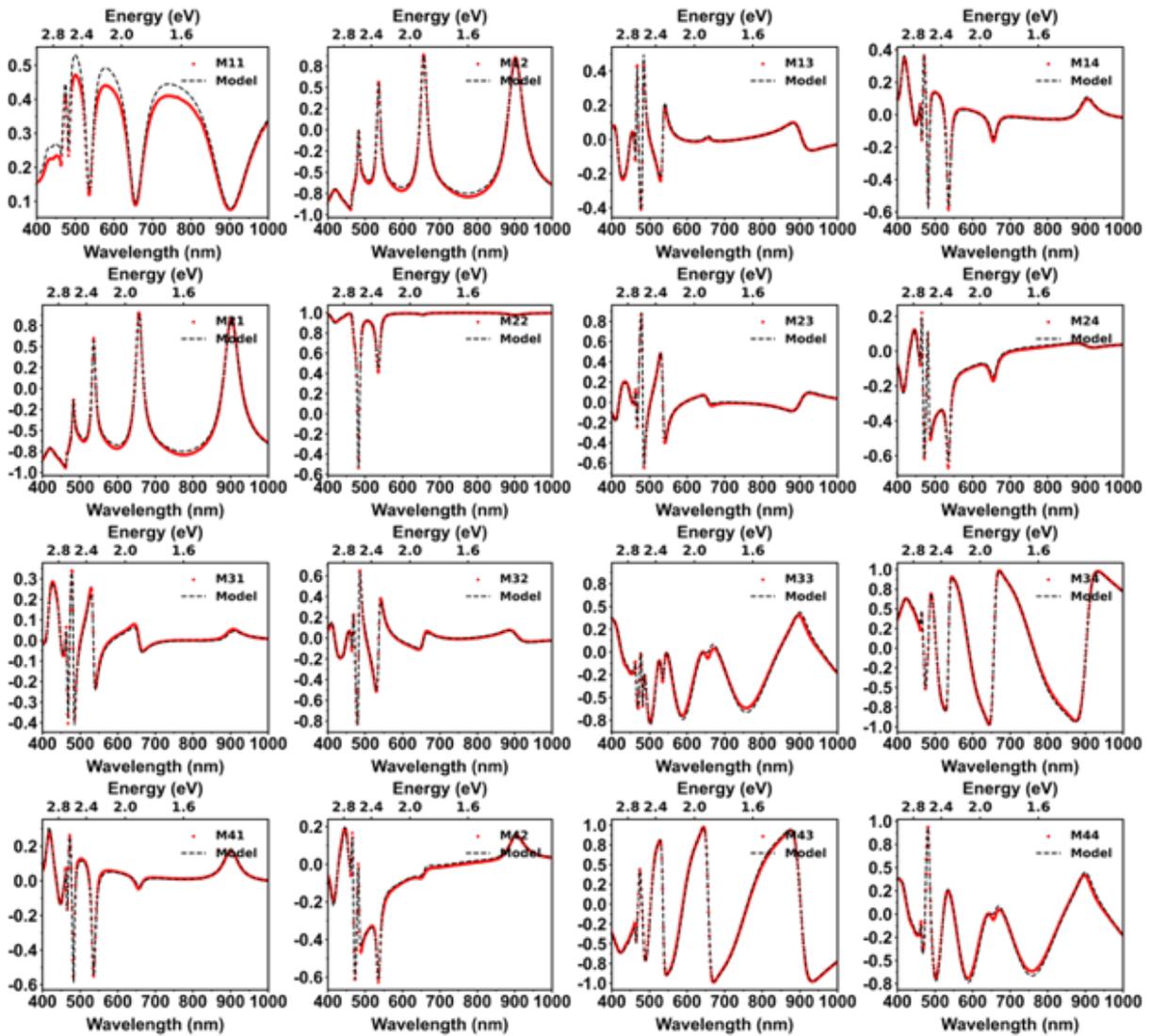

**Figure S6.** Experimental and fitted values of Mueller matrix elements (M11~M44) in the wavelength regime of 400-1000 nm. The two datasets exhibit a high degree of agreement.

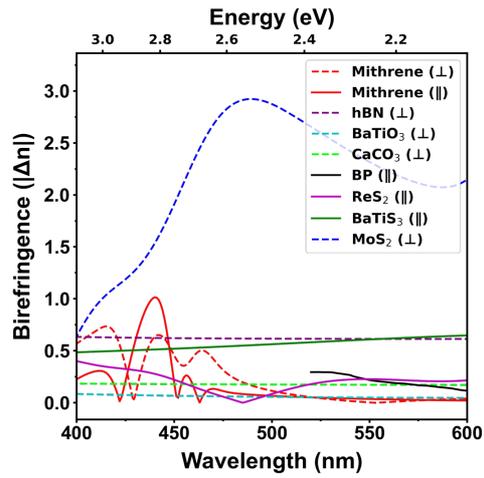

**Figure S7.** A Comparison of the magnitude of in- and out-of-plane birefringence as a function of wavelength with molybdenum disulfide ($MoS_2$) out-of-plane birefringence value[2]. We plot separately for better comparison in Figure 3c.

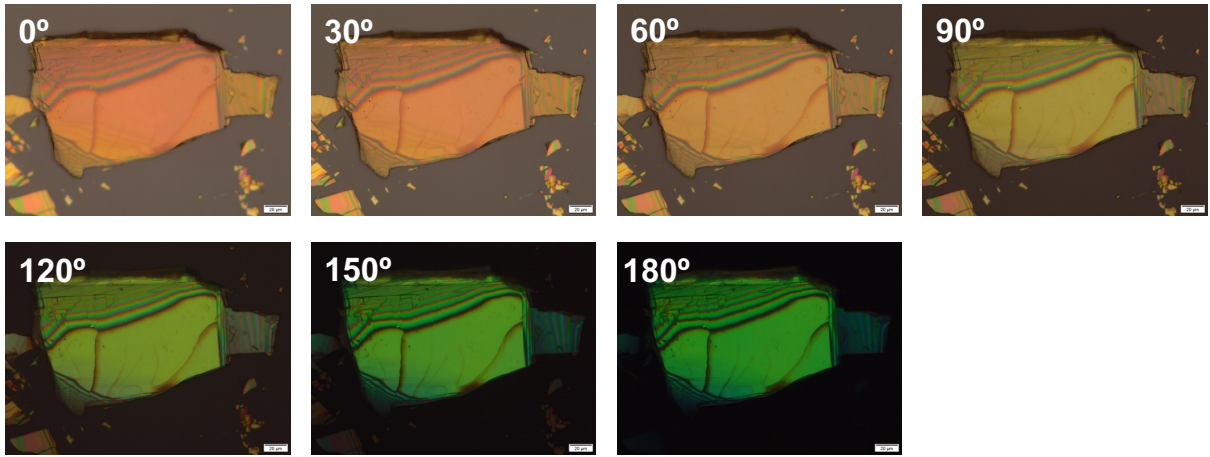

**Figure S8.** Polarized OM image of transferred mithrene as a function of the rotation angle of the polarizer of OM, denoting the stark color difference depending on the rotation angle and in-plane anisotropy.

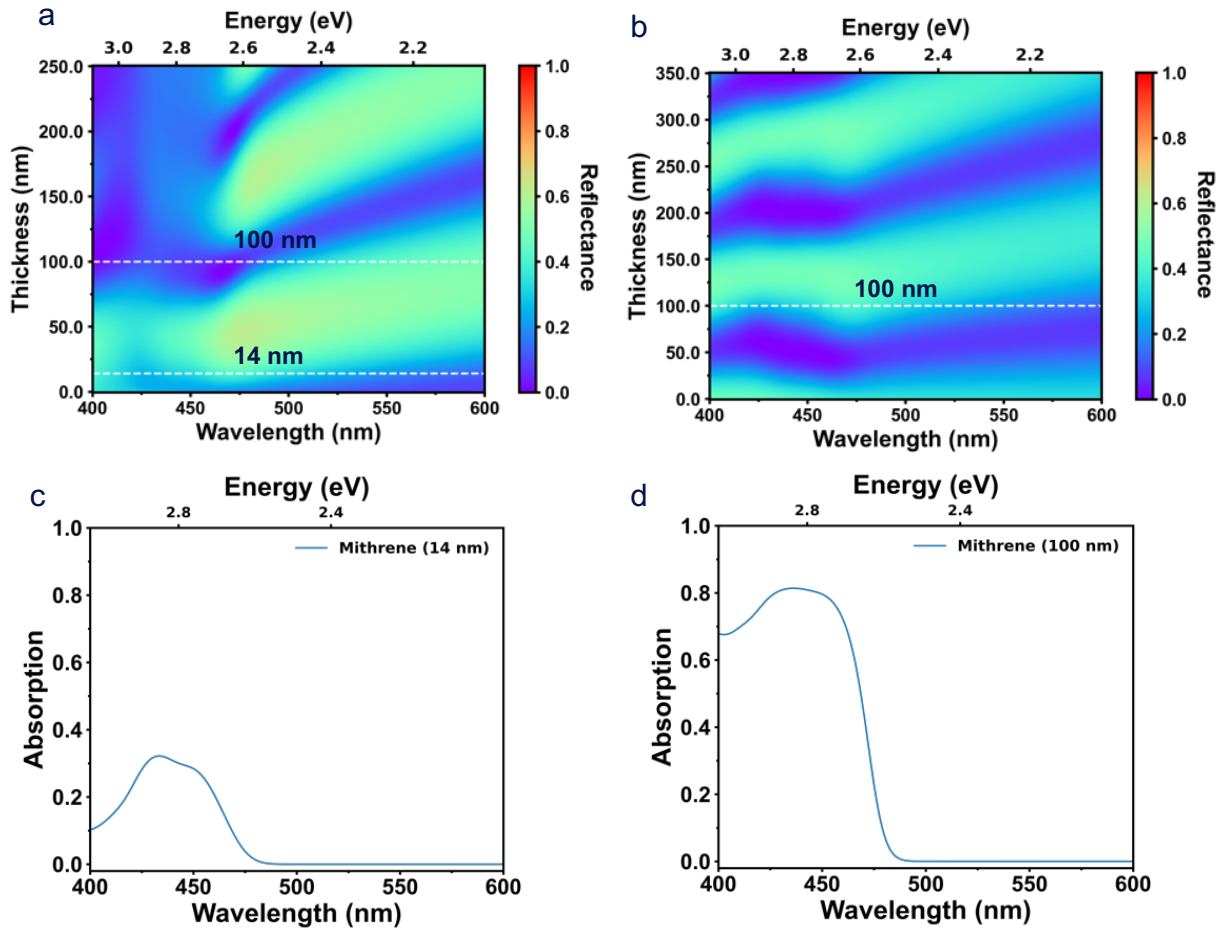

**Figure S9.** The transfer matrix method (TMM) simulation results show the minimized cavity coupling effect between the mithrene and Si/SiO$_2$ substrate. 2D color map of reflectance as a function of the thickness of (a) mithrene (with 100 nm SiO$_2$ beneath) and (b) SiO$_2$ (with 14 nm mithrene above). (c) Absorption spectrum of mithrene (14 nm) / SiO$_2$ (100 nm) / Si, and (d) mithrene (100 nm) / SiO$_2$ (100 nm) / Si.

Using the TMM calculation, firstly, we simulate the reflectance spectrum as a function of mithrene thickness (Figure S9a). The simulation indicates that when the mithrene thickness is 14 nm with 100 nm of SiO2 beneath, mithrene does not show strong absorption. We further calculate the reflectance as a function of SiO$_2$ thickness with a fixed mithrene thickness of 14 nm (Figure S9b). The results also exhibit that the 100 nm thickness of SiO$_2$ is desirable for achieving the minimized cavity effect with a combination of thin mithrene crystals which means that the experimental setup of Figure 3d-f is suitable for the minimized cavity effect. In addition, the absorption spectrum of the mithrene layer is calculated with a "mithrene (14 or 100 nm) / SiO$_2$ (100 nm) / Si" configuration (Figure S9c and d). These results further verify the absence of strong absorption in the thin mithrene (14 nm) layer as shown in Figure S9c unlike the thick mithrene (100 nm) layer (Figure S9d).

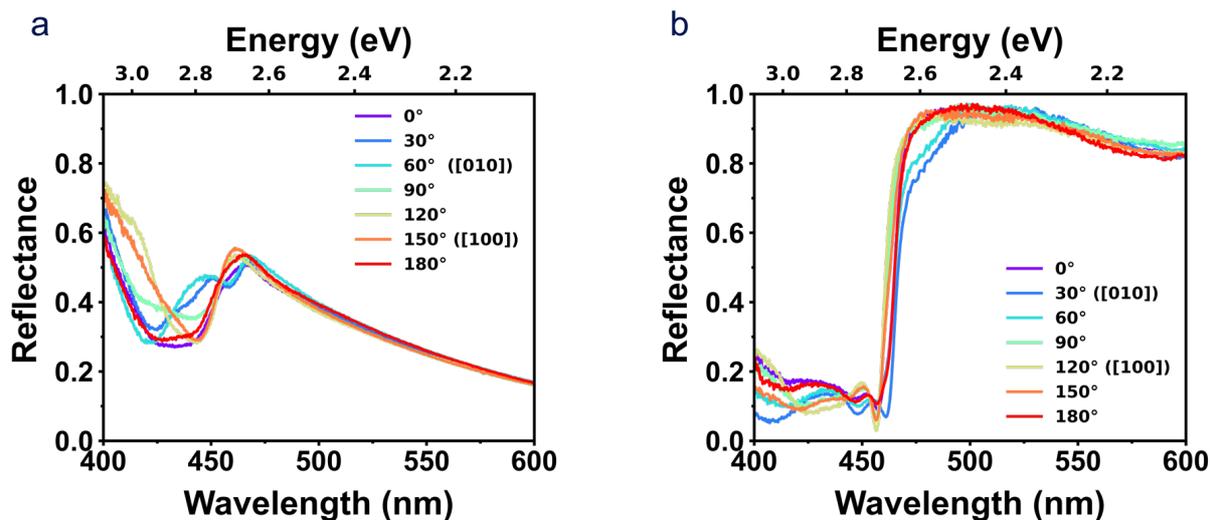

**Figure S10.** The full reflectance spectrum of (a) Figure 3e and (b) Figure 3h depends on the polarization of incidence light.

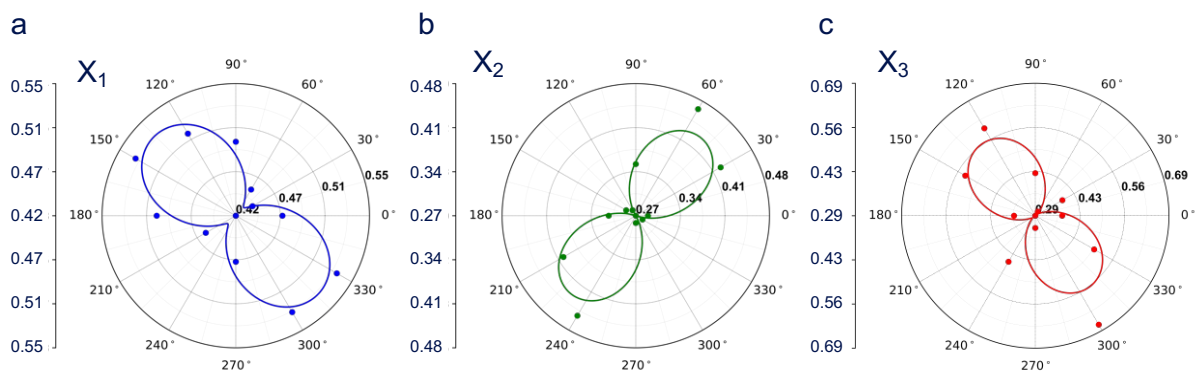

**Figure S11.** The orthogonality between $X_1/X_3$, and $X_2$ excitons (a) The polar plot of $X_1$, (b) $X_2$, and (c) $X_3$ excitons from Figure S10a.

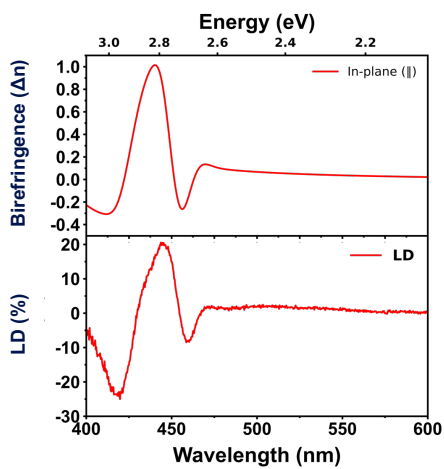

**Figure S12.** A comparison between the birefringence and LD spectrum, showing similar trends since both birefringence and LD are related to the exciton transitions in mithrene.

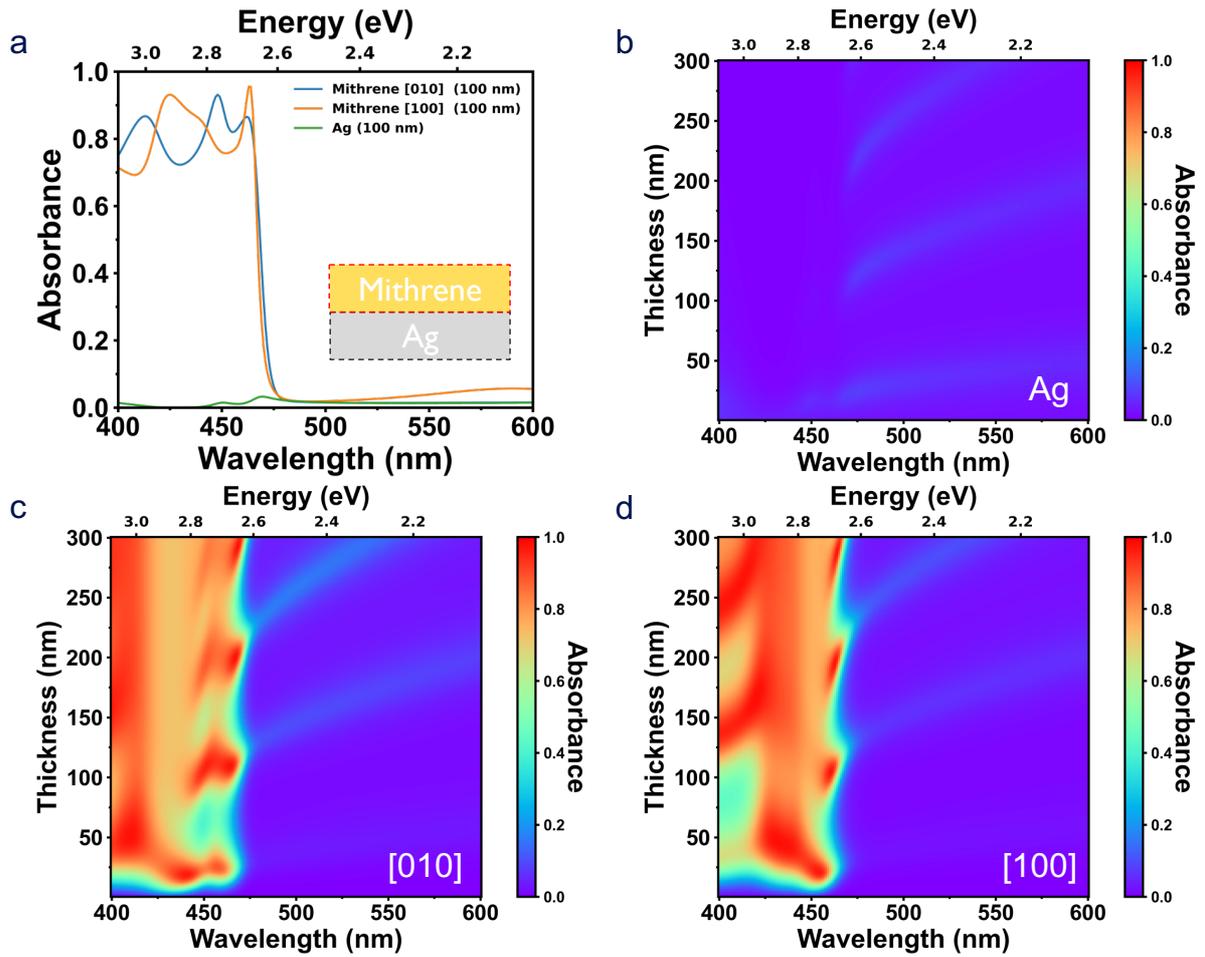

**Figure S13.** (a) The absorption spectrum in the mithrene (100 nm)/Ag (100 nm) structure from each layer. The absorption spectrum as a function of mithrene thickness with the Ag (100 nm) substrate of (b) Ag substrate, (c) mithrene along [010] direction, and (d) mithrene along [100] direction.

First, we simulate the absorption spectrum in the mithrene (100 nm)/Ag (100 nm) structure to distinguish the absorption amount in each layer. As shown in Figure S13a, the Ag substrate exhibits minimal absorption within the visible spectrum, suggesting its efficacy as a reflector within this range. On the other hand, most light is absorbed in the mithrene layer, and the absorption spectrum shows discrepancies along the two optical axes due to its in-plane anisotropic characteristics. We further calculate the absorption of each layer as a function of mithrene thickness with a fixed Ag thickness of 100 nm (Figure S13b). The Ag reflector does not show any significant absorption improvement regardless of the thickness of the mithrene. On the contrary, the mithrene exhibits clearly enhanced absorption whose dispersion follows the polariton branches along two optical axes ([010] and [100]) due to the strong light-matter interactions (Figure S13c and d). Therefore, the improved LD is attributed to mithrene response, not from the Ag substrate.

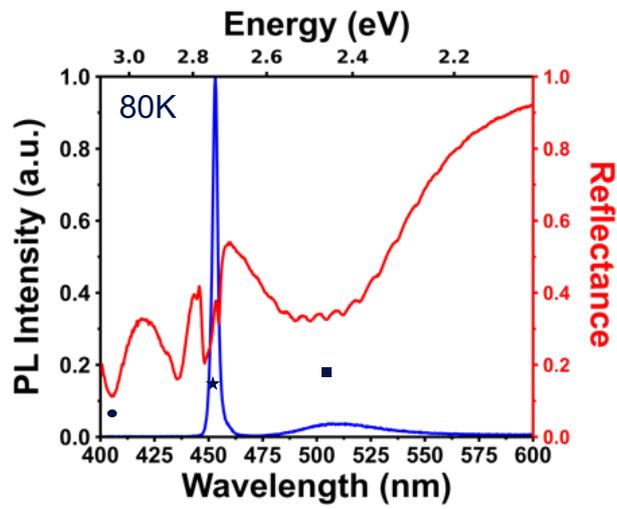

**Figure S14.** The PL spectrum and reflectance spectrum at 80K with unpolarized light incidence shows clear UEP (●), HO (★), and LEP states (■) in the reflectance spectrum. PL spectrum indicates the HO and LEP-induced peak due to self-hybridization.